\documentclass[12pt]{article}

\usepackage[dvips]{graphicx, color}
\begin{document}

\author{Chong-Yu Ruan, Manfred Fink\\ Department of Physics, 
The University of Texas, Austin, Texas 78712}
\date{}
\title{Emission Optics of the Steigerwald Type Electron Gun}
\maketitle

\begin{abstract}
The emission optics of a Steigerwald type electron gun is re-examined. The
virtual and real points of divergence, divergence angles and beam-widths of the electron
beams at different telefocusing strength are measured in detail for first time
. Two different Wehnelt cylinders are used to establish a
contrasting viewpoint. The original `focusing' curves measured by Braucks are 
reconstructed and will be explained only through a `new' interpretation which 
is different from the conventional views. While the image of the emitting surface in front of
the filament is indeed telefocused beyond the anode, the envelope of the beam 
does not `focus' as expected. A new model for the emission mechanism is established based on our results. 
\end{abstract}

\newpage
\section{Introduction}

An electron gun with a telefocus Wehnelt grid was first proposed by K.H. Steigerwald in
1949\cite{Steigerwald1949} ( see Figure \ref{fig:e-gun}). The focusing properties of
this design are based on the shape of the equipotential lines
 inside the Wehnelt cylinder closely imitating a diverging and a converging
lens combination. The name `telefocusing' follows the optical analog. A
detailed experimental verification of the predicted property was
carried out by Braucks\cite{Braucks1958}. By changing the position of 
the inner Wehnelt
cone ($a$)  or the size of the outer Wehnelt opening $\left( d_{w}\right) $,
the strength of the two electron optical lenses will vary with the fields. As shown in Figure \ref{fig:opt},
 Braucks demonstrated the
telefocusing property of the Steigerwald type gun. The focusing
distances $\left( L_{q}\right) $ identified with the smallest beam diameter from the filament tip
were plotted as a function of parameters $%
d_{w}$ and $a$ for his design. Changing the ratio $d_{w}/a$, the focal length could be varied at
 will. An intense electron beam that is naturally collimated without limiting
 apertures makes this gun very attractive for electron diffraction experiment.

Several authors have applied the Steigerwald type gun in gas phase electron 
diffraction units using counting techniques \cite{Fink1970}, \cite{Fink1979}, \cite{Hermann}
 and photographic plates(films)\cite{Taguchi1984}.  To get high count rates and keep the angular 
precision at the same time, it is advantageous that the beam is intense and narrow 
at the intersection of the beam and the gas jet, and the focal point of the beam is set at the 
entrance of the Faraday cage in the detecting plane\cite{Fink78}. In the telefocusing systems,
 the distance between two  electrodes can be varied by the axial movement of the inner
 Wehnelt cone or the anode, thus allowing the focal distance to be adjusted and beam 
quality to be fine tuned. To investigate the effect influenced by the geometry of 
the lensing system,  the beam profile at the proximity of the `focal point' was 
measured and is discussed here for different combinations of $a$, $D$, and $d$.  
It has already been pointed out by Schmoranzer et al.\cite{Schmoranzer1975}, 
and later investigated extensively
 by Schiewe et al.\cite{Schiewe1977} that the sum $a+D$, i.e. the distance from the anode to 
the filament tip, strongly affects the focusing mechanism.
 We built a telefocus gun to probe the optics
 of an electrostatic analyzer. The measured magnification factors of the analyzer
 deviated significantly  from the value one would expect based on computational simulation 
which was verified in several other ways. Two possibilities could account for this:  
the long established telefocusing theory could be wrong, or there could be
an anomalous broadening in the analyzer that is responsible for the differences.
Therefore diagnostic measurements have been setup to re-examine the electron 
beam generated from our telefocus gun, especially to find the focal points 
and the beam profiles in their proximity.

\section{Experiments}

\subsection{Setup}

As depicted in Figure \ref{fig:setup}, the telefocus electron gun was mounted 
on a movable stage 
which was able to move along the viewing line of a Faraday cage,
whose entrance aperture was $50$ micron. Two sets of deflector plates were placed between the
gun and Faraday cage. The first set of deflector plates (aligner) moved along with 
the e-gun so that the emitting direction could
be fine tuned. The second set of deflector plates (scanner) were attached to the detector,
 and a computer
controlled power supply (Bertan 205B) was used to scan the beam
over the aperture to investigate the size of the beam. The
intensity profile of the beam was recorded with a Keithley 616 digital
electrometer connected to a P100 computer through GPIB ports. The scans were
taken at a 15mHz repetition rate. The emission current from the electron gun was set 
to be in the range $\sim 1 \mu A$. The beam profiles, which were always gaussian in
shape, were measured at 5-7 different distances, ranging from $L_{e}=$90mm to 180 mm, 
in order to determine the divergence angle. The size of the beam could be determined in 
two different ways. The full width at half maximum (FWHM) was obtained by fitting 
the profile with a standard gaussian. The FWHM was compared with directly recorded half-maximum points.
The two results agreed with each other within $5\%$. To determine the opening angle of
the electron beam, the FWHMs from different positions were combined to track the envelope. 
However, in some extreme cases the beam profile became very broad that sweeping
 the beam could charge up the insulator behind the deflector plates. Furthermore 
the deflecting field became inhomogeneous, introducing additional error. To avoid this
situation, we estimate the gaussian beam size by taking the inverse of the square 
root of the peak intensity and then
calibrating the results with the cases where reliable scans were available.
Since the peak intensity could be measured very accurately, the above method
was an excellent method for extrapolating gaussian beam for larger beam diameters.
 Two different outer Wehnelt 
cylinders were used to investigate the effects due to field penetration. (A) Wehnelt had
an opening of $d_{w}=3.175cm$, and $d_{w}=1.12cm$ was the value for (B) Wehnelt. 

\subsection{Results and Discussions}

In all cases no focal points have been found. For all the adjustible range of the gun the FWHM of
the beam investigated grew linearly with the distance between the gun and the detector ($L_e$).
3-dimensional plots of the beam-widths measured at different  $d_{w}/a$
 over the range from $L_{e}=0$ to $250mm$ were constructed for two Wehnelt cylinders.
 (Figure \ref{fig:3d_big}, \ref{fig:3d_small})  `Point of divergence' can be defined
 as the point source on the symmetric axis where the beam intersects with
 the zero beam-width plane.  The locations of these `virtual sources' points determine whether 
a real crossover appears along the beam path. Figure \ref{fig:vir_pt} shows that only in
 section marked (c) the real crossover appears in front of the filament; whereas 
in section (b), representing smaller beam-widths, the points of divergence are behind 
the filament. In section (c) the divergent lens contributes most strongly. 
Those crossovers should not be called `focuses' as was recognized first by Braucks 
and later on confirmed by  Schmoranzer et al.\cite{Schmoranzer1975}. The beam-widths in (c) are actually larger 
than those in (b) along the observed distances. It is not clear whether 
a crossover will appear at larger distance as Braucks measured(Figure \ref{fig:opt}). 
We suspect that
Schmoranzer et al. \cite{Schmoranzer1975},\cite{Schiewe1977}
observed cuts at a fixed $L_e$ value and the minimum beam-width was mistaken as `focal point'
 of the beam. Figure \ref{fig:width} shows such cuts measured at $L_{e}=93.2mm$. 
The origin of the coordinate system for $L_{e}$ and the position of the inner Wehnelt cone
 is set at the front edge of the electron gun (Figure 3). By changing the $d_{w}/a$ ratio 
in the gun, the size of the emitting surface and the divergence angle both change. 
The narrowest beam-width observed is the minimization due to the combination of
both factors. In the proximity of the gun, the size factor dominates, but in the far field,
 the dispersion angle is more important. As supported by both Figures \ref{fig:3d_big}
, \ref{fig:3d_small} we found 
 that while moving the detector to larger $L_e$ value, the minimum beam-width starts
from a low $d_{w}/a$ value then gradually moves toward a particular $d_{w}/a$ 
value with the lowest dispersion angle.  This actually agrees with one of the features in 
Braucks' original plot (Figure \ref{fig:opt}) where the $d_{w}/a$ values, corresponding 
to increasing `focal lengths' ($L_q$), converge at one point. To support our argument, 
we inserted into Braucks' plot our measurements, using $L_{q}$ as the distance 
from the filament to the detector, the minimum beam-width was achieved by tuning
$d_{w}/a$. Figure \ref{fig:opt} shows that the $d_{w}/a$ values 
actually agree with
Braucks' measurements. Another feature we reproduced well is that the `density' of the 
function $f(L_{q}, d_{w}/a)$ grows at lower $d_w$. However, the density of $f$ in our measurement 
is much greater than Braucks'. One possible way of explanation may come from $D$, the distance 
from the anode to the Wehnelt cylinder. As pointed out by Schmoranzer and Schiewe, the value of $(D+a)$ 
is a good parameter in the `focusing' mechanism\cite{Schiewe1977}. The $d_{w}/D$ values in our
 setup were larger than those Braucks used. That means our converging field 
is stronger inside the Wehnelt cylinder while the diverging field is weaker.  Since the 
spreading of $L_q$ curves in Braucks' plot was the result  of telefocusing mechanism, 
 it may be reasonable to link the increased density to our stronger convergent fields.

The telefocusing lenses do exist in the Wehnelt cylinder, but a complete ray tracing which 
simulates the beam emission process has never been realized as in other gun types \cite{septier}. The 
general dispersion angle involved in this case is in $mrad$ range, which is difficult to 
simulate; the uncertainty in the initial conditions poses a fundamental problem.
A reverse field created by the self-biasing and the kinetic energy of the thermal
 electrons defines a zero energy surface. This surface, not the filament, serves as the 
emitting object in the imaging system of a telefocusing gun. The shape of this surface depends 
mostly on the local symmetric mechanical design and the electron-electron interaction, but not on the
shape of the filament. This explains why the electron beams are
circular while the hair-pin filament is elongated in one direction. The emission current would
increase exponentially as $a$ decreases however the zero energy surface grows due to a 
greater reverse field created by the self-biasing resistor, and vice versa. 

 In Figure \ref{fig:vir_pt}, the point of divergence appears in front of the filament only for 
small values of $d_{w}/a$ when $a$ is rather large (marked as section c). For very high $d_{w}/a$ (section a),
 the emitting surface is fully  exposed to the anode and no crossover will appear. In the intermediate $d_{w}/a$ range (section b)
 a minimum beam-width might be obtained, but no  
crossover has to occur. This avoids the `Boersch' effect coming from the Coulombic electron-electron
 repulsion at the 
crossover\cite{Zimmerman}, and ensures the fact that the beam divergence angle can be 
kept very small at high emission current. 

As shown in Figure \ref{fig:ang}, it is interesting to 
note that in many cases the relative momentum spread ($\delta p /p$) in phase space, 
which defines the dispersion angle $\alpha$ for each ray bundle emitted from emitting surface, 
is smaller than the thermal broadening, $\alpha_{T}=\sqrt{{kT}/eV} \sim 4.18 \times 10^{-3}$ rad. 
Particularly for Wehnelt cylinder (A), the average transverse velocity of the thermal electrons is 
reduced by a factor of 10 at the minimum beam-width condition. 
Generally in a simplified treatment one assumes that the transverse velocity distribution changes only
little in the emission process, however this is only true for electrons leaving the charge cloud perpendicularly.
In the transverse direction, the equipotential lines of the local reverse field act as filters and reduce the emittance angles. 
It also indicates that the angular spread of the beam 
envelope is mostly from the inherent thermal broadening than from the lensing
 process in the gun. Therefore by using filament with lower 
operating temperature, e.g. barium oxide, hexaboride cathodes, beams with smaller angular spread can be obtained. It is noteworthy that in section (b) marked in Figure \ref{fig:vir_pt}, the electron would 
acquire highest brightness since both beam-width and dispersion angle were minimized.


The answer to why the beam envelope created by a telefocuing electron gun
does not converge actually can be illustrated also by its optical analog. In a
telescope, the remote object is imaged upon the eye through a
telefocusing mechanism. The envelope of all the incoming light rays, traced back, is diverging as 
it will ultimately reach the star we are observing. Similarly for an electron gun, it should be 
that the image of the emitting surface be reconstructed remotely while the beam envelope as a whole 
keeps diverging. As will be discussed in more details in the following article, the very same 
electron gun with (B) Wehnelt is used to probe the imaging optics of a spherical analyzer. Two
beam-profiles are measured very accurately at the entrance and exit of the analyzer. Based on the 
measurements of the beam envelope obtained in this work, and the ratio of the two beam profiles,
 the location of the images could be deconvoluted. The position of the image of the emitting 
surface at different $d_{w}/a$ is shown in Figure \ref{fig:img} where indeed the source to image 
distance ($L_{q}$) increases as the strength of the telefocusing increases($d_{w}/a$ decreases).

We propose a new model to describe the performance of a telefocus electron 
gun which is illustrated in Figure \ref{fig:illus}. In case (a) the inner Wehnelt cylinder is most retracted from the 
anode, the total emission is very low, and the reverse biasing field is rather weak. The emitting 
surface thus is very close to the filament and small; also a real crossover is observed to form in the gun. 
These two features facilitate a small beam size at the proximity of the gun, as shown in Figure \ref{fig:3d_big}. 
This geometry also creates the strongest diverging fields which account for the largest imaging
 length measured, as shown in Figure \ref{fig:img}. Moving the inner cylinder further out, the real crossover withdraws 
toward the emitting surface. In case (b) the electron gun is on the verge of having a real 
crossover inside the Wehnelt cylinder. While the total emission current is increasing, the emitting surface begins 
to move out of the filament
and grows in size. As the inner Wehnelt cylinder continues to be moved forward, the strength of the converging and 
diverging fields adjust accordingly.  In case (c) the electron gun has the least spreading beam and correspondingly the 
furthermost virtual point of divergence. Total emission current grows further, and the larger emitting 
surface contributes to a wider beam-width in the proximity of the gun compared to case (a) but produces the narrowest 
beam in the far field due to its very small angular spread. No real crossover is present anywhere in the emission process. 
This relieves the energy broadening from 
coulombic repulsion force. As the inner Wehnelt cylinder is further moved forward, in case (d) , the emitting 
surface is greatly exposed to the anode voltage; the very weak diverging fields account for the shortest imaging 
length. While the emission current is the greatest, the beam-width and the dispersion angle 
are also greatest in this geometry. Note that in case (c) the minimum beam-width condition is 
actually the interplay of the diverging and converging lenses inside the Wehnelt cylinder. Surprisingly, by reducing the 
self-biasing resistor and hence increasing the total emission current, the lensing fields do not change much. 
Also the growth of the emitting surface is insignificant as verified by a small increase in the beam-width. 
This weak response is caused by the very great potential gradient in the proximity of the emitting surface which
downgrades the reverse biasing fields in this situation. The lensing fields obviously operate in larger scale and depend 
most critically on geometrical factors. 
Note that the spreading of the beam is highly exaggerated in Figure \ref{fig:illus} to help illustrate the emission profiles. In reality, the overall beam envelopes and the singular 
emitting profiles coming from point sources on the emitting surface very much overlap with each other.

\section{Conclusion}
The very small dispersion angle found in (A) Wehnelt cylinder demonstrated
the superb ability of telefocus gun to create a very narrow, well collimated 
and intense electron beam with very simple design. Our new model explains the changes in the emission optics 
of the Steigerwald type gun from generating focused beams to creating parallel beams. 
Throughout the investigation, the emission current of the electron gun was set to 
$\sim 0.1 \mu A$ range to avoid complications due to Coulomb repulsion force at the emitting surface. 
It was found that the positions of points of convergence stayed within 10\% while 
the current could be increased up to a factor of 100. Also our $d_{w}/a$ values in Figure \ref{fig:opt}
 agreed well to Braucks' results although the general structures of the electron guns were
 different. These two facts point to the possibility of studying the emission optics of the
 telefocusing electron gun systematically based on geometric parameters (e.g. $d_{w}/a$, $d_{w}/D$) in 
spite of the fact that the emission surface can not be constructed reliably. 

\section{Acknowedgement}
One of the authors wishes to express his gratitude to Dominik Hammer for helpful discussions, and 
to Chris Zernial for making the graphs of the emission optics. This work was supported by Texas
 Advanced Research Project and Robert A. Welch Foundation.

\newpage

\begin{figure}[htbp]
  \begin{center}
    \caption{The Steigerwald type electron gun.(A)Filament, (B)Self-biasing resistors, (C)Outer Wehnelt cylinder, (D)The inner Wehnelt cone, (E)Anode.}
    \label{fig:e-gun}
  \end{center}
\end{figure}
\begin{figure}[htbp]
  \begin{center}
    \caption{The lines show the reconstruction of the $L_{q}$ (`focal distance') curve based on Braucks' paper. The points have been extracted from our measurements.}
     \label{fig:opt}
  \end{center}
\end{figure}
\begin{figure}[htbp]
  \begin{center}    
    \caption{Experimental apparatus: (A)electron gun; (B) and (C) deflector plates, (D) Faraday cage and aperture.}
    \label{fig:setup}
  \end{center}
\end{figure}
\begin{figure}[htbp]
  \begin{center}
    \caption{The 3D plot of the beam-widths over the distance and different $d_{w}/a$ for Wehnelt cylinder (A).}
    \label{fig:3d_big}
  \end{center}
\end{figure}
\begin{figure}[htbp]
  \begin{center}
        \caption{The 3D plot of the beam-widths over the distance and different $d_{w}/a$ for Wehnelt cylinder(B).}
    \label{fig:3d_small}
  \end{center}
\end{figure}
\begin{figure}[htbp]
  \begin{center}
    \caption{The point of divergence. Note that crossovers are formed when the points of divergences rise above the inner Wehnelt cone position (slanted line).}
    \label{fig:vir_pt}
  \end{center}
\end{figure}
\begin{figure}[htbp]
  \begin{center}
    \caption{Beam-widths(FWHM) at $L_{e}=93.2mm$}
    \label{fig:width}
  \end{center}
\end{figure}
\begin{figure}[htbp]
  \begin{center}
    \caption{Beam dispersion angle.}
    \label{fig:ang}
  \end{center}
\end{figure}
\begin{figure}[htbp]
  \begin{center}
    \caption{The image positions as determined from the data with a spherical analyzer.}
    \label{fig:img}
  \end{center}
\end{figure}
\begin{figure}[htbp]
  \begin{center}
    \caption{The electron beams emitted from a telefocus electron gun}
    \label{fig:illus}
  \end{center}
\end{figure}

\end{document}